\documentclass{article}

\usepackage{cite}
\usepackage{titlesec}
\usepackage{floatrow}
\usepackage{graphicx}
\usepackage[label font=bf]{subfig}
\usepackage{caption}
\usepackage{pdfpages}
\usepackage{multicol}        
\usepackage[bottom]{footmisc}
\usepackage{subfig}
\usepackage{amsfonts}
\usepackage{hyperref}
\usepackage[cmex10]{amsmath}
\usepackage{amsmath}
\usepackage{amsthm}
\usepackage{xcolor}
\usepackage{mathtools}
\usepackage{amssymb}

\newfloatcommand{capbtabbox}{table}[][\FBwidth]

\theoremstyle{remark}

\usepackage{bm}

\usepackage[normalem]{ulem}
\usepackage{xcolor}

\newcommand{\add}[1]{\textcolor{black}{#1}}
\newcommand{\del}[1]{\textcolor{red}{}}

\usepackage{geometry}

\usepackage{blindtext}

\begin{document}
	\floatsetup[figure]{style=plain,subcapbesideposition=top}
	\title{Constant-selection evolutionary dynamics on weighted networks}
	\author{Jnanajyoti Bhaumik$^1$ \and Naoki Masuda$^{1,2,3}$}
	\date{%
		$^1$Department of Mathematics, State University of New York at Buffalo, NY 14260-2900, USA\\%
		$^2$ Institute for Artificial Intelligence and Data Science, University at Buffalo, State University of New York at Buffalo,
			Buffalo, NY 14260-5030, USA\\%
		$^3$ Center for Computational Social Science, Kobe University, Kobe, 657-8501, Japan\\[2ex]
	}
	
	\maketitle

\begin{abstract}
%
%
The population structure often impacts evolutionary dynamics. In constant-selection evolutionary dynamics between two types, amplifiers of selection are networks that promote the fitter mutant to take over the entire population, and  suppressors of selection do the opposite. It has been shown that most undirected and unweighted networks are amplifiers of selection under a common updating rule and initial condition. Here, we extensively investigate how edge weights influence selection on undirected networks. We show that random edge weights make small networks less amplifying than the corresponding unweighted networks in a majority of cases and also make them suppressors of selection (i.e., less amplifying than the complete graph, or equivalently, the Moran process) in many cases. Qualitatively, the same result holds true for larger empirical networks. These results suggest that amplifiers of selection are not as common for weighted networks as for unweighted counterparts.

\end{abstract}
\section{Introduction}

Evolutionary dynamics models enable us to study how populations change over time under natural selection. Although effects of population structure on evolutionary dynamics have been studied for decades, a seminal paper by Lieberman and colleagues spurred mathematical and numerical studies of evolutionary dynamics under any population structure modeled as networks, which are often referred as evolutionary graph theory~\cite{lieberman2005evolutionary}. One of the simplest setting of the evolutionary graph theory is to consider dynamics in which two types, which are called resident and mutant types and have a type-dependent constant fitness value, stochastically compete on the given network \cite{lieberman2005evolutionary,nowak_book,antal2006evolutionary,nowak2010evolutionary,broom2011evolutionary}. We call this dynamics the constant-selection dynamics. An individual of either type is assumed to occupy a node of the network, and individuals with the larger fitness reproduce on the neighboring nodes with a higher frequency. The constant-selection dynamics model allows us to systematically compute, among other things, the probability and time to fixation, i.e., the situation in which all the nodes are eventually monopolized by one type. In the absence of mutation, which we assume throughout the present paper as many other papers do, fixation is the absorbing state of the evolutionary dynamics, and how fixation occurs crucially depends on the network structure.

\begin{figure}[t] 
	\begin{center}
		\includegraphics[width=8cm]{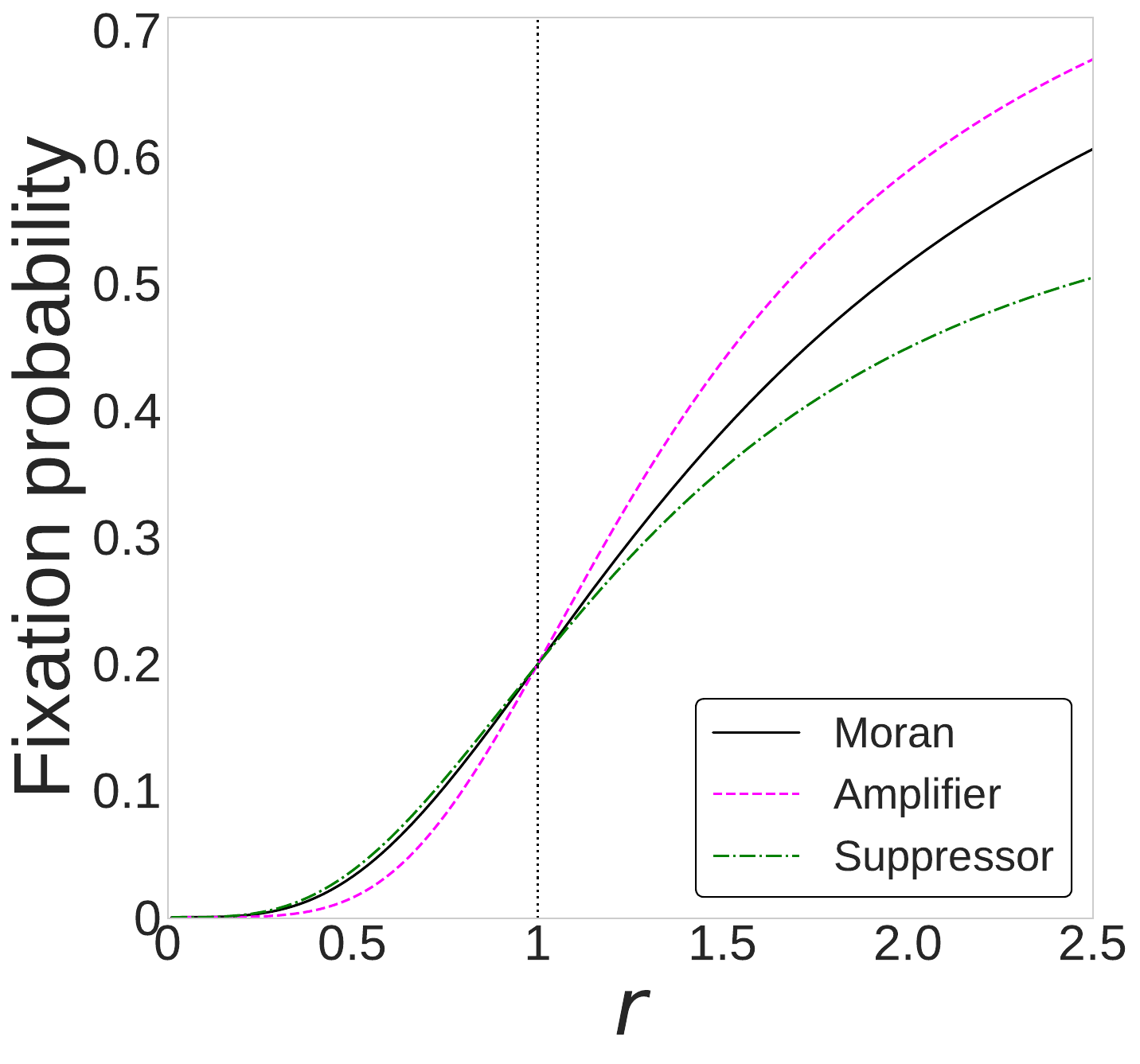}
		\caption{Visualization of amplifiers and suppressors of selection relative to the Moran process. The different lines represent the fixation probability as a function of the fitness of the mutant, $r$. We set the fitness of the resident to $1$. The vertical dotted line marks $r=1$, at which the mutant and resident types have the same strength and the fixation probability is equal to $1/N$ for all the three curves.	}
		\label{fig:amp_sup_schematic_figures}
	\end{center}
\end{figure}

In constant-selection dynamics, there are amplifiers of selection, which are defined as networks that amplify the fitness difference of the fitter type. In other words, the fitter type is more likely to fixate in an amplifying network compared to in the Moran process (i.e., the complete graph). In contrast, suppressors of selection suppress the fitness difference in the sense that the fitter type is less likely to fixate than in the Moran process. We schematically illustrate an amplifier of selection, suppressor of selection, and the Moran process in Fig.~\ref{fig:amp_sup_schematic_figures}. In fact, most of the undirected unweighted networks are known to be amplifiers of selection under the most common combination of the updating rule (i.e., the birth-death rule, introduced in section~\ref{sec:model}) and initial condition (i.e., uniform initialization, or uniform distribution of the single initial mutant over the nodes in the network).
For example, 100 out of 112 (89.3\%)  networks on six nodes, 791 out of 853 (92.7\%) networks on seven nodes, and 10,544 out of 11,117 (94.8\%) networks on eight nodes are amplifiers of selection~\cite{alcalde2018evolutionary}. In contrast, suppressors of selection are rare under the same condition; 1 out of 112 (0.89\%) networks on six nodes, 3 out of 853 (0.35\%) networks on seven nodes, and 90 out of 11,117 (0.81\%) networks on eight nodes are suppressors of selection~\cite{hindersin2015most,alcalde2017suppressors,alcalde2018evolutionary}.
However, research has shown that this prevailing result that most networks are amplifiers of selection is not robust against generalizations of the network model, even if one pertains to the birth-death rule combined with uniform initialization. Specifically, amplifiers of selection are substantially less common and suppressors of selection are more common for small directed networks~\cite{masuda2009directionality}, small temporal (i.e., time-varying) networks~\cite{Bhaumik2023switching},
hypergraphs of various sizes~\cite{ruodan2023hypergraph}, and multi-layer networks of various sizes~\cite{ruodan2023multilayer}. In the present study, we ask whether amplifiers or suppressors of selection are common for weighted networks, which are a most basic extension of the simple (i.e., undirected and unweighted) networks. 

Weighted networks may substantially change constant-selection evolutionary dynamics on networks. For example,
edge weights can be exploited for constructing various amplifiers of selection \cite{svoboda2024amplifiers, adlam2015amplifiers, pavlogiannis2018construction}. Furthermore, weighted networks can provide substantial amplification with a negligible increase in the fixation time~\cite{tkadlec2019population,tkadlec2021fast}. Introduction of edge weights also facilitates generation of amplifiers of selection under the so-called death-birth updating~\cite{tkadlec2020limits,allen2020transient}, whereas these amplifiers are mathematically known to be of ``transient'' type~\cite{tkadlec2020limits}. Furthermore, the fan graph, proposed in~\cite{allen2020transient}, changes from a suppressor of weak selection to an amplifier of weak selection under the birth-death updating rule as a single edge weight increases~\cite{allen2021fixation}.
%

Despite these studies, however, relationships between edge weights and the strength of selection remain elusive. In this paper, we provide multiple lines of evidence to suggest the abundance of suppressors of selection among weighted networks. For simplicity, we assume that the network is undirected.
Our main finding is that suppressors of selection are much more common for weighted networks than their unweighted counterparts. We also find that one can also systematically create either amplifiers or suppressors of selection by varying the edge weights, as previous studies showed \cite{allen2020transient, allen2021fixation}, in the case of weighted networks with high symmetry.

\section{Model}\label{sec:model}

Consider a static undirected weighted network with $N$ nodes. At every discrete time step, we assume that each node is either a resident or mutant. The resident and mutant have fitness $1$ and $r$, respectively. The fitness represents the propensity with which each type is selected for reproduction in each time step. The mutant type is assumed to initially occupy just one node, which is selected uniformly at random among the $N$ nodes. The other $N-1$ nodes are initially occupied by the resident type.

We then run the Birth-death (Bd) process, which is a generalization of the Moran process to networks \cite{lieberman2005evolutionary, ohtsuki2006simple, olfati2007evolutionary, nowak2010evolutionary,shakarian2012review,perc2013evolutionary}. Specifically, in every discrete time step, we select a node $v$ to reproduce with the probability proportional to its fitness value. Next, we select a neighbor of $v$, denoted by $v'$, with the probability proportional to the weight of the undirected edge ($v$, $v'$). Then, the type at $v$ (i.e., either resident or mutant) replaces that at $v'$. We repeat this process until the entire population is of a single type, either resident or mutant, which we call the fixation. 

We also study the death-Birth (dB) updating rule~\cite{kaveh2015duality,hindersin2015most,pattni2015evolutionary,tkadlec2020limits,allen2020transient}. With the dB rule, one first selects a node $v$ to die uniformly at random in each time step. Then, one selects a neighbor $v'$ of $v$ with the probability proportional to the product of $v'$'s fitness and the weight of edge ($v$, $v'$). The type at $v'$ replaces that of $v$ to complete a single time step.

\section{Methods for computing the fixation probability}

As has been done in other studies, we investigate the fixation probability of a single initial mutant.
We compute the fixation probability of weighted networks as follows \cite{lieberman2005evolutionary,hindersin2016exact}. The state of the constant-selection evolutionary dynamics on the network is specified by binary vector $\bm{s}=\left( s_1,\ldots,s_N\right)^{\top}$, where $s_i=0$ or $s_i=1$ if the $i$th node is inhabited by a resident or mutant, respectively, and $^{\top}$ represents the transposition. There are $2^N$ states. To describe the transition probability matrix $T$ among the states, we use the $N\times N$ weighted adjacency matrix of the network, denoted by $W = (W_{ij})$, where $W_{ij}$ is the weight of edge ($i$, $j$). To delineate the states, we require an indexing order. We achieve this using a bijective function $f$ from the set of the states, denoted by $\bm{S}$, to $\{1, \ldots, 2^N\}$. Let $\bm{s}$ denote a state with $m$ mutants, where $s_i = 1$ for $i \in \{g(1), \ldots, g(m)\}$ and $s_i = 0$ for $i \in \{g(m+1), \ldots, g(N)\}$, with $g$ being a permutation of $\{1, \ldots, N\}$. 

Consider state $\bm{s'}$ having $m+1$ mutants, such that $s'_i = 1$ for $i \in \{g(1), \ldots, g(m), g(\del{m+1}\add{\ell})\}$ and $s'_i = 0$ for $i \in \{g(m+\add{1}\del{2}), \ldots,\add{g(\ell-1),g(\ell+1),\ldots,} g(N)\}$. Here, the difference between $\bm{s}$ and $\bm{s'}$ is solely at position $g(\del{m+1}\add{\ell})$, corresponding to a transition of a resident in $\bm{s}$ to a mutant in $\bm{s'}$. The probability of the transition from $\bm{s}$ to $\bm{s'}$ is given by
\begin{equation}
	T_{f(\bm{s}),f(\bm{s'})} = \frac{r}{rm + N-m}\sum_{m'=1}^{m}\frac{W_{g(m'), g(\del{m+1}\add{\ell})}}{\overline{w}(g(m'))},
\end{equation}
where $\overline{w}(i)$ is the strength (i.e., weighted degree) of the $i$th node and equal to $\sum_{j=1}^N W_{ij}$.

Consider another state $\bm{s''}$ with $m-1$ mutants, where $s_i'' = 1$ for $i \in \{ g(1), \ldots, g(\tilde{m}-1), g(\tilde{m}+1), ..., g(m)\}$ and $s''_i = 0$ for $i \in \{ g(\tilde{m}), g(m+1), \ldots, g(N) \}$, with $\tilde{m}\in \{ 1, \ldots, m \}$.
Then, the probability of the transition from $\bm{s}$ to $\bm{s''}$ is given by
\begin{equation}
	T_{f(\bm{s}),f(\bm{s''})} = \frac{1}{rm + N-m}\sum_{m'=m+1}^{N}\frac{W_{g(m'),g(\tilde{m})}}{\overline{w}(g(m'))}.
\end{equation}

Finally, the probability that state $\bm{s}$ does not change in one time step is equal to
\begin{equation}
	T_{f(\bm{s}), f(\del{\bm{s''}}\add{\bm{s}})} =
	1 - \frac{1}{rm+N-m}\left[\sum_{\ell=m+1}^{N}\sum_{m'=1}^{m}\frac{W_{g(m'),g({\ell)}}}{\overline{w}(g(m'))}
	+\sum_{\tilde{m}=1}^{m}\sum_{m'={m+1}}^{N}\frac{W_{g(m'),g(\tilde{m})}}{\overline{w}(g(m'))}\right].
\end{equation}

We let $x_{f\left(\bm{s}\right)}$ be the probability of fixation when the dynamics start from $\bm{s}$. We observe that
\begin{equation}
\label{eqn:recursive-fixation}
	x_{f(\bm{s})}=\sum_{\bm{s'}\in S}T_{f(\bm{s}),f(\bm{s'})}x_{f(\bm{s'})}.
\end{equation}
Let $\bm{x}$ denote the vector of length $2^N$ that contains $x_{f\left(\bm{s}\right)}$ for all states $\bm{s} \in \bm{S}$.
Then, we can succincctly write Eq.~\eqref{eqn:recursive-fixation} as follows:
\begin{equation}
	\bm{x}=T\bm{x}.
\end{equation}
Note that $T$ is a $2^N \times 2^N$ matrix. Because $x_{f\left( \left[0,\ldots,0\right] \right)}=0$ and $x_{f\left( \left[1,\ldots,1\right] \right)}=1$, we only need to solve a set of $2^N-2$ linear equations. The fixation probability $\rho$ under the uniform initialization, i.e., when the location of the single initial mutant is selected from all the nodes uniformly at random, is given by
\begin{equation}
	\rho = \frac{1}{N}\sum_{\bm{s'}\in \bm{S_1}}x_{f(\bm{s'})},
\label{eq:rho-uniform-final}	
\end{equation}
where $\bm{S_1}$ denotes the set containing the $N$ states having just one mutant.

We define amplifiers and suppressors of selection as follows; similar definitions were used in the literature \cite{lieberman2005evolutionary,voorhees2013birth}. Let $\rho_{G}\left(r\right)$ and $\rho_{M}\left(r\right)$ denote the fixation probability of a network $G$ and of the Moran process (i.e., Bd process on the complete graph) with the same number of nodes, respectively, when the fitness of the mutant is equal to $r$.
It is known that (see e.g.\,\cite{nowak_book})
\begin{equation}
	\rho_M(r) = \frac{1-\frac{1}{r}}{1-\frac{1}{r^{N}}}.
	\label{eq:fixation-prob-Moran}
\end{equation}
If $\rho_{G}\left(r\right)<\rho_{M}\left(r\right)$ when $r<1$ and $\rho_{G}\left(r\right)>\rho_{M}\left(r\right)$ when $r>1$, then  network $G$ is an amplifier of selection. If $\rho_{G}\left(r\right)>\rho_{M}\left(r\right)$ when $r<1$ and $\rho_{G}\left(r\right)<\rho_{M}\left(r\right)$ when $r>1$, then $G$ is a suppressor of selection. To convey the concept of amplifier and suppressor of selection, we compare $\rho_{G}\left( r \right)$ for a hypothetical amplifier of selection, $\rho_{G}\left( r \right)$ for a hypothetical suppressor of selection, and $\rho_{M}\left(r\right)$ in Fig.~\ref{fig:amp_sup_schematic_figures}.

For a given network $G$, we only numerically computed the fixation probability at several values of $r$. We say that the network is an amplifier of selection if $\rho_{G}\left(r\right) < \rho_{M}\left(r\right)$ at $r \in \{ 0.7,0.8,0.9 \}$ and $\rho_{G}\left(r\right) > \rho_{M}\left(r\right)$ at $r \in \{ 1.1,1.2,1.3,1.4,1.5,1.6 \}$ and that it is a suppressor of selection if $\rho_{G}\left(r\right) > \rho_{M}\left(r\right)$ at $r \in \{ 0.7,0.8,0.9 \}$ and $\rho_{G}\left(r\right) < \rho_{M}\left(r\right)$ at $r \in \{ 1.1,1.2,1.3,1.4,1.5,1.6 \}$. Some networks are neither amplifier nor suppressor of selection \cite{alcalde2018evolutionary}.
\add{It should be noted that, similar to a majority of work on constant-selection dynamics, we do not assume weak selection, which would correspond to $r$ only slightly different from $1$.}

\section{Results}\label{sec:Results}

	\subsection{Networks on six nodes}\label{sec:SmallNetworks}
	
We first analyzed the fixation probability in connected networks with six nodes. There are 112 non-isomorphic undirected connected networks on six nodes. \add{The rationale behind the analysis of six-node networks is that they allow us to compute the analytical solutions owing to a feasible number of equations (i.e., $2^6 - 2 = 62$) without assuming particular symmetry in the network structure.} For each non-isomorphic network, we assigned each edge in the network a random weight independently distributed according to the uniform density on $(0, 1]$, generating a weighted network. \add{Empirical weighted networks have various weight values, and the uniform density is a simple model of such variability.} We generated 100 such weighted networks for each of the 112 networks on six nodes. We analyzed the fixation probability for each weighted network under the Bd updating rule and uniform initialization and classified it into either amplifier of selection, suppressor of selection, or neither. We also analyzed the fixation probability under the dB updating rule in the same manner. Under dB updating rule, the only amplifiers of selection were transient amplifiers (i.e., their amplification effect disappears as fitness increases beyond a certain threshold) even if one allows directed and weighted networks \cite{tkadlec2020limits}.

We show the number of amplifiers, suppressors, neither networks in Table~\ref{table:stats}. The table indicates that 3862 out of 11200 (34.5\%) of the weighted networks are suppressors of selection under the Bd rule. This result is in stark contrast with that for unweighted networks on six nodes, for which only one of the 112 networks is a suppressor of selection~\cite{alcalde2017suppressors}.
Table~\ref{table:stats} also indicates that
there is no amplifier and that most of the networks (99.3\%) are suppressors of selection under the dB rule.
This result is consistent with that for unweighted networks~\cite{hindersin2015most, alcalde2017suppressors}.

\begin{figure}[t] 
		\begin{center}
		\includegraphics[width=13cm]{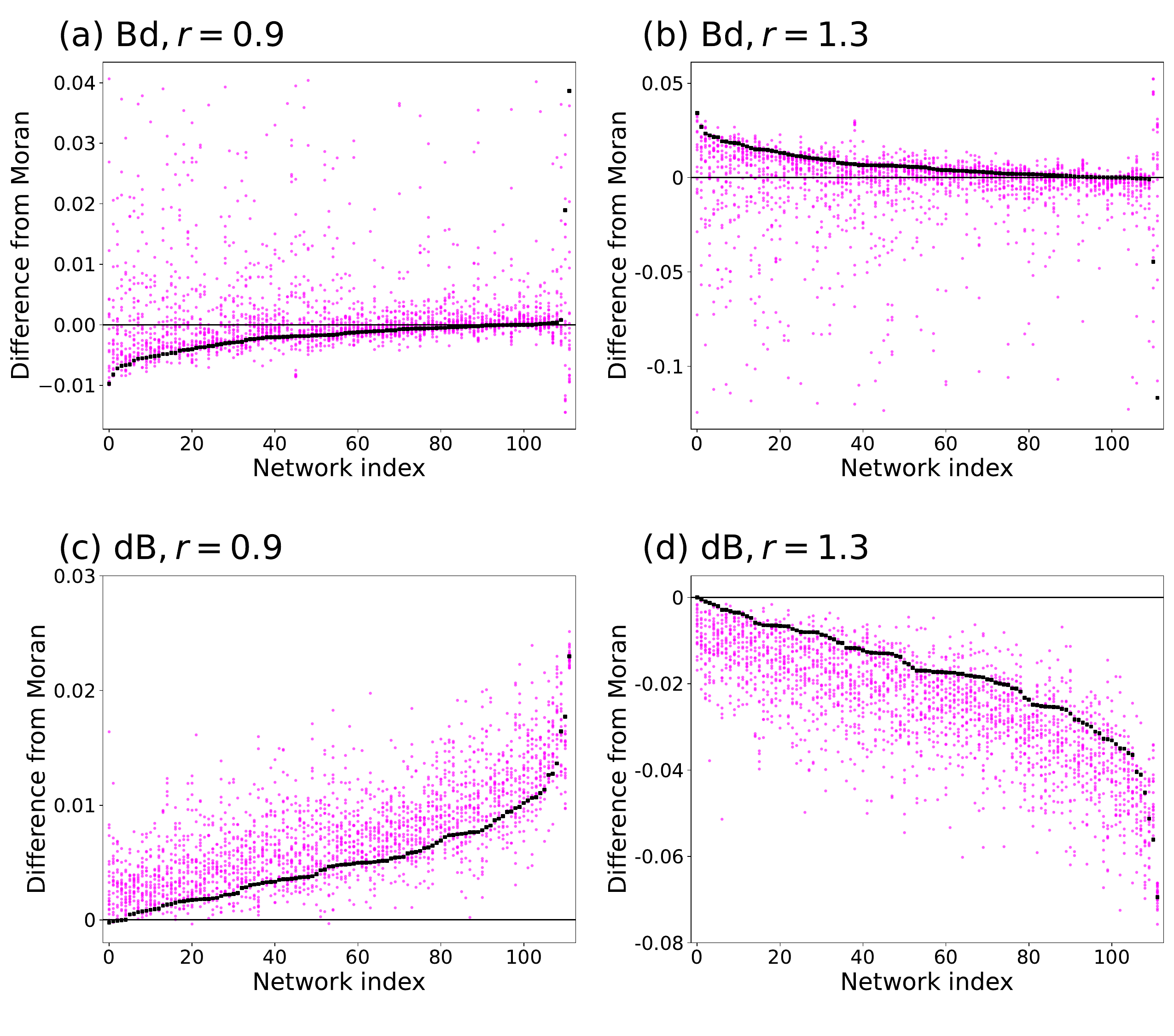}
		\caption{Fixation probability of weighted networks on six nodes, for the Bd and dB updating. The horizontal axis represents the index of the 112 non-isomorphic unweighted networks on six nodes. For each unweighted network, we assigned random edge weights to generate a weighted network 20 times. The vertical axis represents the difference from the Moran process in terms of the fixation probability. (a) Bd rule, $r=0.9$. (b) Bd rule, $r=1.3$. (c) dB rule, $r=0.9$. (d) dB rule, $r=1.3$. \add{A magenta circle represents a weighted network. A black square represents the corresponding unweighted network.}
		}
		\label{fig:bdscatter}
	\end{center}
\end{figure}

To further compare between the unweighted and weighted networks, we computed the difference in the fixation probability between any six-node weighted network and the Moran process at two values of $r$, i.e., $r=0.9$ and $1.3$. If the difference is negative at $r=0.9$ and positive at $r=1.3$, it is strongly suggested that the weighted network is an amplifier of selection. In contrast, if the difference is positive at $r=0.9$ and negative at $r=1.3$, the network is likely to be a suppressor of selection. For the Bd rule, we plot the thus computed difference in the fixation probability at $r=0.9$ and $r=1.3$ in Figs.~\ref{fig:bdscatter}(a) and \ref{fig:bdscatter}(b), respectively. The horizontal axis represents the index of the network. We indexed the 112 networks in ascending order of the fixation probability of the unweighted network in Fig.~\ref{fig:bdscatter}(a) and in descending order of the fixation probability of the unweighted network in Fig.~\ref{fig:bdscatter}(b). The small squares in black represent the fixation probability of the unweighted networks. It should be noted that the indexing order of the networks is different between Fig.~\ref{fig:bdscatter}(a) and Fig.~\ref{fig:bdscatter}(b), whereas the two orderings tend to be similar. For each of the 112 unweighted networks on six nodes, we randomly generated 20 weighted networks according to the same procedure as those used in Table~\ref{table:stats} and calculated the difference in the fixation probability from the Moran process. The small circles in \del{blue} \add{magenta} represent the difference for each weighted network from the Moran process in terms of the fixation probability.

Figure~\ref{fig:bdscatter} shows that a large portion of the weighted networks
have a positive difference in the fixation probability relative to the Moran process (i.e., a positive value on the vertical axis) at $r=0.9$ (45.0\%; see Fig.~\ref{fig:bdscatter}(a)) and a negative difference
at $r=1.3$ (43.8\%; see Fig.~\ref{fig:bdscatter}(b)). 
We have confirmed that 43.1\% of the weighted networks satisfy both of the these two properties and therefore are strongly suggested to be suppressors of selection.
As expected, this last fraction is consistent with the numbers reported in Table~\ref{table:stats}. 
Furthermore, the figure shows that a large portion of the weighted networks 
have a higher fixation probability than the corresponding unweighted network at $r=0.9$
(74.3\%; see Fig.~\ref{fig:bdscatter}(a)) and
a lower fixation probability than the unweighted network at $r=1.3$
(75.7\%; see Fig.~\ref{fig:bdscatter}(b)).
Therefore, adding random edge weights to an unweighted network is more likely to cause the network to become more suppressing than vice versa.
	
We plot the corresponding results for the dB updating rule in Figs.~\ref{fig:bdscatter}(c) and \ref{fig:bdscatter}(d). 
Consistent with the results shown in Table~\ref{table:stats}, these figures, showing few negative values in Fig.~\ref{fig:bdscatter}(c) and positive values in Fig.~\ref{fig:bdscatter}(d), affirm that few of these weighted networks are amplifiers of selection. Furthermore, as in the case of the Bd updating rule,
adding random edge weights is likely to make the network more suppressing under the dB updating rule.
Specifically,
83.0\%  of the weighted networks 
have a higher fixation probability than the corresponding unweighted network at $r=0.9$
(see Fig.~\ref{fig:bdscatter}(c)), and
85.3\% of the weighted networks have a lower fixation probability than the unweighted network at $r=1.3$
(see Fig.~\ref{fig:bdscatter}(d)).
Therefore, we conclude that adding edge weights to six-node networks strongly tends to make the network more suppressing under both Bd and dB rules.
	
	\begin{table}[t] 
		\centering
		\begin{tabular}{cccc} 
			\hline
			Rule & Amplifier & Suppressor  & Neither\\ \hline
			Bd& 6165& 3862& 1173\\
			dB& 0&  11125 &  75\\ \hline
			\hline
		\end{tabular}
		{\caption{Number of amplifiers and suppressors of selection among $11200$ connected undirected and weighted networks on six nodes.}
			\label{table:stats}}
	\end{table}

\subsection{Larger symmetric networks}\label{sec:symmetric}
	
In this section, we examine the fixation probability of larger weighted networks with symmetry, i.e., weighted complete and star graphs.  \add{The symmetry of the networks allows us to analytically compute the fixation probability of much larger networks than six-node networks because the fixation probability of the nodes in the symmetric position (technically called structurally equivalent nodes) have the same fixation probability. To be able to exploit the symmetry of the network, one cannot assign random edge weights independently for all edges. Therefore, we have decided to construct minimal models of edge weight with which we systematically vary the edge weight $w$ shared by a particular group of nodes.} In this and the following sections, we focus on the Bd updating rule.
	
\subsubsection{Weighted complete graphs}\label{sec:completegraph}

\begin{figure}[t] 
		\begin{center}
			\includegraphics[width=13cm]{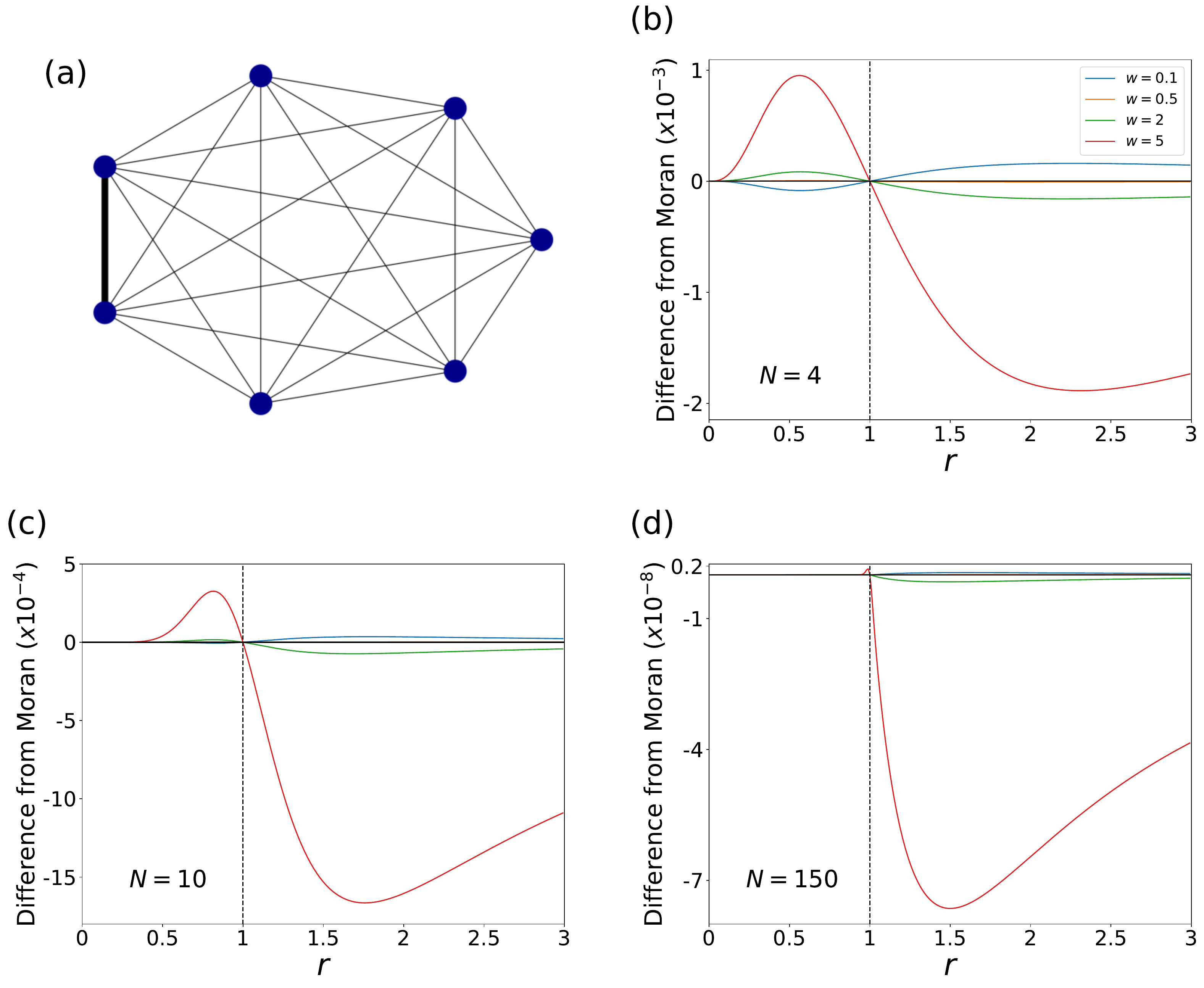}
			\caption{Fixation probability of weighted complete graphs with a single edge with a different edge weight. (a) An example network with $N=8$ nodes. The single edge with weight $w \neq 1$ is shown by the thick line. (b)--(d) Fixation probability relative to that of the Moran process for the weighted complete graphs of the type shown in (a). (b) $N=4$. (c) $N=10$. (d) $N=150$. In (b)--(d), the dotted lines represent $r=1$.}
	\label{fig:complete_graph_edge_weight_2}
\end{center}
\end{figure}

The complete graph is isothermal, i.e., the fixation dynamics on it is trivially equivalent to the Moran process.
We consider weighted complete graphs that have one edge of weight $w$ and the remaining edges with edge weight $1$.
See Fig.~\ref{fig:complete_graph_edge_weight_2}(a) for an example with $N=8$. Because the two nodes connected with the edge with weight $w$ are structurally equivalent to each other, so are the other $N-2$ nodes, we only need to solve a set of linear equations with $3(N-1)$ unknowns to obtain the fixation probability, $\rho$ (see \ref{sec:appendix_A-transition-probabilities-for-the-weighted-complete-graph}). Figures~\ref{fig:complete_graph_edge_weight_2}(b), (c), and (d) show the difference between the weighted complete graph with $N =  4, 10$, and $150$ nodes, respectively, and the Moran process in terms of $\rho$. Figure~\ref{fig:complete_graph_edge_weight_2}(b) indicates that the weighted complete graph with $N=4$ nodes is a suppressor of selection when $w= 0.5$, $2$, and $5$ because the difference is positive for $r<1$ and negative for $r>1$. We note that $\rho$ of the weighted complete graph is visually close to that of the Moran process when $w=0.5$; the orange line in Figure~\ref{fig:complete_graph_edge_weight_2}(b) is almost hidden behind the black horizontal line marking $0$.
In contrast, the weighted complete graph with $N=4$ is an amplifier of selection when $w=0.1$ because the difference is negative for $r<1$ and positive for $r>1$. Additionally, as $w$ increases, the weighted complete graph becomes a stronger suppressor of selection. The results for $N=10$ and $N=150$, shown in Figs.~\ref{fig:complete_graph_edge_weight_2}(c) and (d), respectively, are qualitatively the same as those for $N=4$. The suppressing effect becomes stronger as $w$ increases and weaker as $N$ increases.

\begin{figure}[!htbp]
	\begin{center}
		\includegraphics[width=13cm]{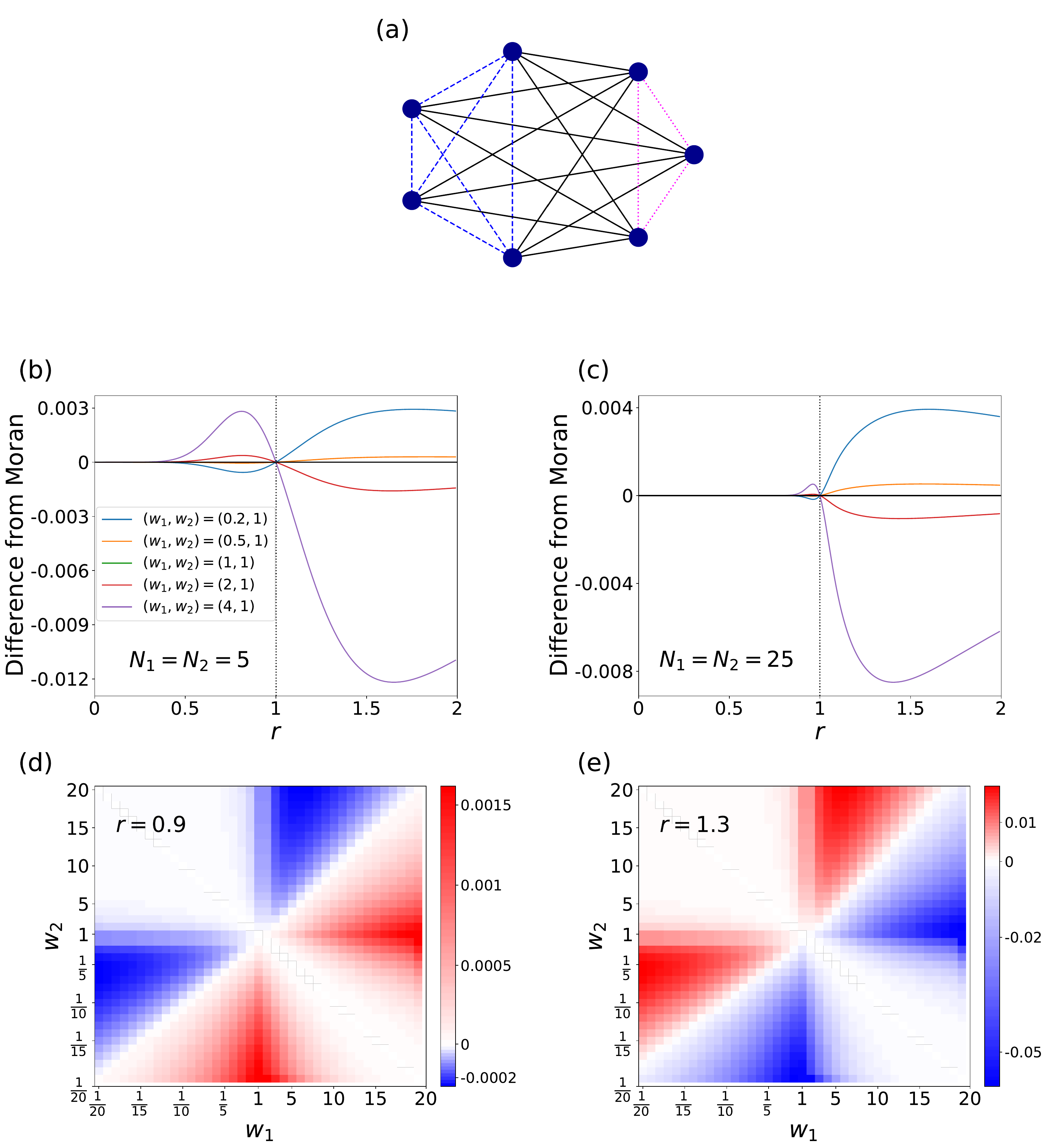}
		\caption{Fixation probability of weighted complete graphs with two larger groups of nodes. (a) An example network with $N_1 = 4$ and $N_2 = 3$. The edges with weight $1$, $w_1$, and $w_2$ are shown in solid, dashed, and dotted lines, respectively. (b) and (c): Fixation probability relative to that of the Moran process for the weighted complete graphs of the type shown in (a);
the dotted lines represent $r=1$.
We set $N_1=N_2=5$ in (b) and $N_1=N_2=25$ in (c). (d) and (e): Fixation probability of the same type of weighted complete graphs with $N_1=N_2=50$ and varying $w_1$ and $w_2$.
(d) $r=0.9$. (e) $r=1.3$.}
		\label{fig:complete_graph_w1_w2}
	\end{center}
\end{figure}

To assess the generality of these results, we consider a wider family of weighted complete graphs constructed as follows. We divide the nodes in a complete graph into two sets, one with $N_1$ nodes, and the other with $N-N_1 \equiv N_2$ nodes. We set the weights of the edges between pairs of the $N_1$ nodes to $w_1$, and those for the edges between pairs of the $N_2$ nodes to $w_2$. The weight of the edges connecting a node in the first set and a node in the second set is $1$. See Fig.~\ref{fig:complete_graph_w1_w2}(a) for an example with $N_1 = 4$ and $N_2 = 3$.
The weighted complete graphs analyzed in Fig.~\ref{fig:complete_graph_edge_weight_2} are a special case of the present family of weighted complete graphs with $N_1 = 2$, $w_1 = w$, and $w_2 = 1$. Because of the structural equivalence among the $N_1$ nodes and that among the $N_2$ nodes, we only need to solve a set of linear equations with $\left(N_1+1\right)\times\left(N_2+1\right)$ unknowns to obtain the fixation probability at each value of $r$. We describe the set of linear equations, including the special case considered with Fig.~\ref{fig:complete_graph_edge_weight_2}, in \ref{sec:appendix_A-transition-probabilities-for-the-weighted-complete-graph}.

In Figs.~\ref{fig:complete_graph_w1_w2}(b) and (c), we show the fixation probability relative to that for the Moran process in the case of $N_1=N_2=5$ (therefore, $N=10$) and $N_1=N_2=25$ (therefore, $N=50$), respectively. We set $w_2=1$ and vary $w_1$. The figure indicates that, for both values of $N$,
the weighted complete graphs are amplifiers of selection when $w_1<1$ and suppressors of selection when $w_1>1$. The networks are stronger suppressors of selection when $w_1$ is larger. 
These results are similar to those shown in Figs.~\ref{fig:complete_graph_edge_weight_2}(b)--(d) except that,
in Figs.~\ref{fig:complete_graph_edge_weight_2}(b)--(d), $w=0.5$ yields a weak suppressor of selection, whereas
the same value of $w$ yields an amplifier of selection in Figs.~\ref{fig:complete_graph_w1_w2}(b) and (c).
Note that the unweighted complete graph, corresponding to $w_1 = 1$, is isothermal, which is consistent with  Figs.~\ref{fig:complete_graph_w1_w2}(b) and (c). 

In Figs.~\ref{fig:complete_graph_w1_w2}(d) and (e), we show the fixation probability of the weighted complete graphs with $N_1=N_2=50$ relative to that of the Moran process, as a function of $w_1$ and $w_2$. We set $r=0.9$ and $r=1.3$ in
Figs.~\ref{fig:complete_graph_w1_w2}(d) and (e), respectively. In both Figs.~\ref{fig:complete_graph_w1_w2}(d) and (e), the weighted complete graphs and the Moran process have the same fixation probability when $w_1 = w_2$, shown in white. This result is expected because the weighted complete graph is an isothermal graph when $N_1=N_2$ and $w_1=w_2$. Furthermore, the weighted complete graph is an amplifier of selection (i.e., region shown in blue in 
Fig.~\ref{fig:complete_graph_w1_w2}(d) and red in Fig.~\ref{fig:complete_graph_w1_w2}(e)) roughly when
$w_1 w_2 > 1$. Conversely, it is a suppressor of selection (i.e., region shown in red in 
Fig.~\ref{fig:complete_graph_w1_w2}(d) and blue in Fig.~\ref{fig:complete_graph_w1_w2}(e)) roughly when $w_1 w_2 < 1$.

\subsubsection{Weighted star graphs}\label{sec:stargraph}

\begin{figure}[t]
	\begin{center}
		\includegraphics[width=15.5cm]{	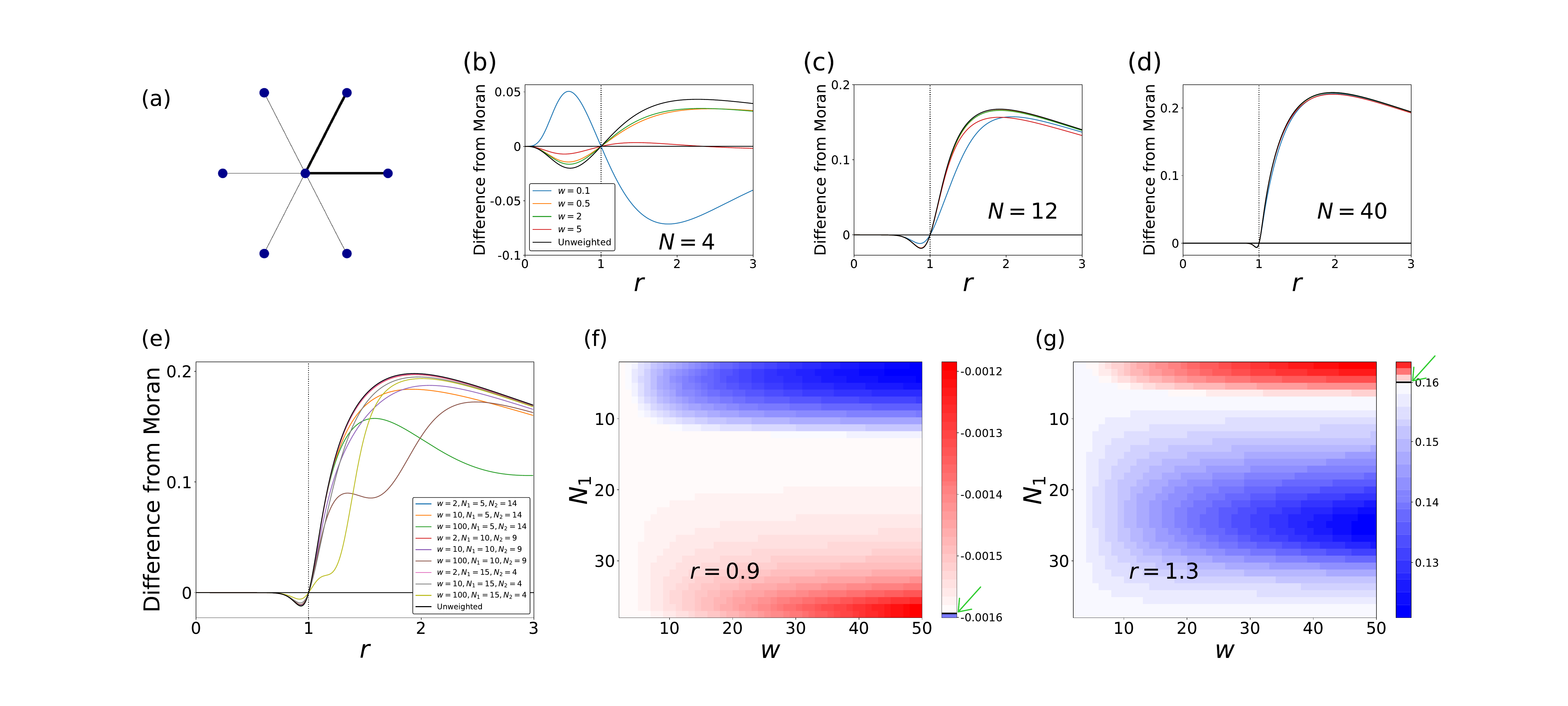}
		\caption{
%
Fixation probability of weighted star graphs. (a) A weighted star graph with $N_1 = 2$ and $N_2 = 4$. The edges with weight $w$ are shown in thick lines. \add{(b) Fixation probability relative to that of the Moran process for the weighted star graphs with $N=4$ and $N_1=1$ as $w$ varies. (c) Same for $N=12$ and $N_1=3$. (d) Same for $N=40$ and $N_1 = 10$.} (\del{b}\add{e}) Fixation probability relative to that of the Moran process for various weighted star graphs on 20 nodes. \add{In (b)--(e), the} \del{The} dotted  line\add{s} represent\del{s} $r=1$. (\del{c}\add{f}) and (\del{d}\add{g}): Fixation probability of weighted star graphs with $N=40$ and various $N_1$ and $w$ values. (\del{c}\add{f}) $r=0.9$. (\del{d}\add{g}) $r=1.3$. Note that $1 \le N_1 \le N-2 = 38$.
In (\del{c}\add{f}) and (\del{d}\add{g}), the \add{black} horizontal lines and the \add{green} arrows pointing to them in the color bar show the values for the unweighted star graph.}
		\label{fig:weightedstar}
	\end{center}
\end{figure}

Unweighted star graphs are known to be strong amplifiers of selection under the combination of the Bd rule and the uniformly distributed initial mutant~\cite{lieberman2005evolutionary,pavlogiannis2018construction}.
In this section, we consider weighted star graphs in which we divide the $N$ nodes into three sets, one node set $V_1$ with $N_1$ leaf nodes, another node set $V_2$ with $N_2$ different leaf nodes, and a single hub node, where $N_1 + N_2 + 1 = N$. There are $N-1$ leaf nodes in total. Each leaf node in $V_1$ is adjacent to the hub node by an edge with weight $w$. Each leaf node in $V_2$ is adjacent to the hub node by an edge with weight $1$. 
See Fig.~\ref{fig:weightedstar}(a) for an example.
We derive the transition probabilities of the fixation dynamics on the weighted star graph in \ref{sec:appendix_B-transition-probabilities-for-the-weighted-star-graph}. We need to solve a set of linear equations with $2 (N_1+1)(N_2+1)$ unknowns.
 
\add{In Figs.~\ref{fig:weightedstar}(b), (c), and (d), we show the fixation probability, $\rho$, of the weighted star graph relative to that for the Moran process when $N=4$, $12$, and $40$, respectively, each with four values of $w$. We set $N_1 = N/4$ in these figures. Figure~\ref{fig:weightedstar}(b) shows that the weighted star graph with $N=4$ nodes (and therefore $N_1 = 1$) is a suppressor of selection when $w=0.1$.
The weighted star graph with $N=4$ is a weaker amplifier of selection than the corresponding unweighted star graph when $w=0.5$ and $w=2$. Last, it is a transient amplifier of selection~\cite{allen2020transient} when $w=5$. In other words, it transitions from being an amplifier of selection to being a suppressor of selection approximately at $r = 2.35$ as $r$ increases. Overall, the weighted complete graph with $N=4$ nodes is more suppressing than the unweighted counterpart across these values of $w$. The weighted star graphs with $N=12$ (see Fig.~\ref{fig:weightedstar}(c)) and $N=40$ (see Fig.~\ref{fig:weightedstar}(d)) nodes are also more suppressing than their unweighted counterparts.}

Figure~\ref{fig:weightedstar}\del{(b)}\add{(e)} shows \del{the fixation probability,} $\rho$\del{,} of the weighted star graph relative to that for the Moran process when $N=20$, with three values of $N_1$ and three values of $w$. Note that we set $w>1$ without loss of generality because the network remains the same if one swaps $V_1$ and $V_2$ and changes $w$ to $1/w$. The figure indicates that all the weighted star graphs are less amplifying than the unweighted star graph. In Figs.~\ref{fig:weightedstar}\del{(c)}\add{(f)} and \del{(d)}\add{(g)}, we plot the difference between the weighted star graph and Moran process in terms of $\rho$ when $N=40$, for different values of $N_1$ (and hence $N_2 = N - N_1-1$) and $w$. We set $r=0.9$ in Fig.~\ref{fig:weightedstar}\del{(c)}\add{(f)} and $r=1.3$ in Fig.~\ref{fig:weightedstar}\del{(d)}\add{(g)}. We have marked the value of $\rho$ for the unweighted star graph using a black line on the color bar (at $-0.00160$ in Fig.~\ref{fig:weightedstar}\del{(c)}\add{(f)} and $0.160$ in Fig.~\ref{fig:weightedstar}\del{(d)}\add{(g)}, indicated by arrows) and made it correspond to white on the color scale. We find that a large portion of the parameter space makes the weighted star graphs weaker amplifiers of selection than the unweighted star graph (i.e., region shown in red in Fig.~\ref{fig:weightedstar}\del{(c)}\add{(f)} and blue in Fig.~\ref{fig:weightedstar}\del{(d)}\add{(g)}; roughly for the combination of any $w$ and $N_1\geq 11$). In this parameter region, a larger value of $w$ makes the weighted star graph less amplifying, and this result is consistent with those shown in Fig.~\ref{fig:weightedstar}\del{(b)}\add{(e)}. In contrast, Figs.~\ref{fig:weightedstar}\del{(c)}\add{(f)} and \del{(d)}\add{(g)} suggest that the weighted star graph with $N_1 \le 7$ is a stronger amplifier of selection than the unweighted star graph. 

\add{To summarize the results for the symmetric networks, we find that it is possible to deliberately assign the edge weights to easily make the weighted network more suppressing than the corresponding unweighted network. }

\subsection{Empirical networks}\label{sec:Empirical-networks}
	
In this section, we numerically simulate the Bd process on \del{the following} six empirical weighted networks. \add{We provide detailed descriptions of the empirical networks in \ref{sec:description-of-the-empirical-networks}.}
\add{The rationale behind this analysis is that many empirical networks are weighted networks. Here we compare empirical weighted networks against the case in which the edge weight is ignored. Another purpose of the present analysis is to validate our findings for the six-node networks and symmetrical networks presented in the previous sections on the empirical networks, which are larger than six-node networks and asymmetric.} 

\del{First, we use a social network of
 raccoons}\del{~\cite{reynolds2015raccoon}.} \del{The data were collected using proximity logging collars on a wild suburban raccoon population within a 20-ha area of Ned Brown Forest Preserve in Cook County, Illinois, USA, observed for $52$ weeks from July 2004 to June 2005. An event was recorded whenever two raccoons came in close proximity (i.e., 1-1.5m) of each other. A node represents a raccoon. An edge represents a proximity event between two raccoons. We set the edge weight to the total time that the given pair of raccoons spent in proximity. In total, there are $N=24$ nodes and $1997$ edges.}
\del{\del{Second, we chose a primate network based on a wild group of $N=25$ \textit{Macaca fuscata} individuals in Yakushima, Japan} \cite{griffin2012community}\del{. The population was free ranging and not captive. An edge represents grooming interaction between the individuals. There are 1340 edges. An event implies an interaction for greater than one minute. The edge weight is equal to the number of grooming interactions between the two individuals.}}
\del{\del{Third, we use data from an ants' colony}~\cite{ants2015quevillonsocial}\del{. Each of the $N=39$ nodes represents an ant in a colony. Each of the 330 edges represents a trophallaxis event, which was recorded when the two ants were engaged in mandible-to-mandible contact for greater than one second. The edge weight is equal to the number of events involving the two ants.}\del{Fourth, we use a social network of sparrows }\cite{arnberg2015social}\del{. A flock was defined as a group of birds within an approximately 5-meter radius. A node represents a bird. Edges represent the so-called simple ratio association index}~\cite{cairns1987comparison} \del{distributed between 0 and 1, encoding flock co-membership. There are $N=40$ nodes and 305 edges.}}\del{\del{Fifth, we use contacts between members of five households in the Matsangoni sub-location within the Kilifi Health and Demographic Surveillance Site (KHDSS) in coastal Kenya}~\cite{kilifi2016quantifying}\del{. A household was characterized as a collection of individuals who shared a common kitchen. Each participant in the study wore a sensor capable of detecting another sensor within a 1.5-meter radius. Each node represents a household member. An edge denotes a recorded interaction between two members. The study documented 47 nodes and 219 pairwise interactions across individuals from different households and 32,426 pairwise interactions within the same household. The edge weight is equal to the number of interaction events between two individuals.}\del{Sixth, we use a human contact network in a hospital}~\cite{hospitalvanhems2013estimating}\del{. Data collection took place in a university hospital's geriatric unit in Lyon, France, between December 6, 2010, at 1 pm, and December 10, 2010, at 2 pm. Nineteen beds were located in the unit. Thirty-one patients were hospitalized, and 50 professionals worked in the unit during the recording period. Among a total of 81 individuals, $N=75$ individuals, i.e., 29 patients and 46 medical staff, participated in the study. Among the medical staff were $27$ nurses or nurses’ aides, $11$ medical doctors, and $8$ administrative staff members. An edge represents a time-stamped contact between two individuals; there are $32,424$ time-stamped edges. The edge weight is the number of times an event occurred between a pair of nodes.}\del{We acquired these networks from \url{https://networkrepository.com/}~[nr-aaai15].} \del{We use the Bd updating rule in this section.}} To initialize each simulation, we place a mutant on a node selected uniformly at random. Then, in each time step, we select a node, denoted by $v$, that reproduces with the probability proportional to the fitness. Then, we select a neighbor of $v$ for death uniformly at random. We repeat this process until all the nodes were of the same type. For each network and  value of $r$, we carried out $3.5 \times 10^5$ simulations in parallel on $40$ cores, giving us a total of $120 \times 10^5$ simulations. We obtained the fixation probability as the fraction of runs in which the mutant fixated. We simulated the weighted networks with $r \in \{0.7,0.8,0.9,1,1.1,1.2,1.3,1.4,1.5,1.6\}$ for all the networks \del{except the hospital network of $75$ nodes. For the hospital network, we omitted $r=1.5$ and $1.6$ due to high computational cost}.
	
\begin{figure}[!htbp]
\begin{center}
	\includegraphics[width=15cm]{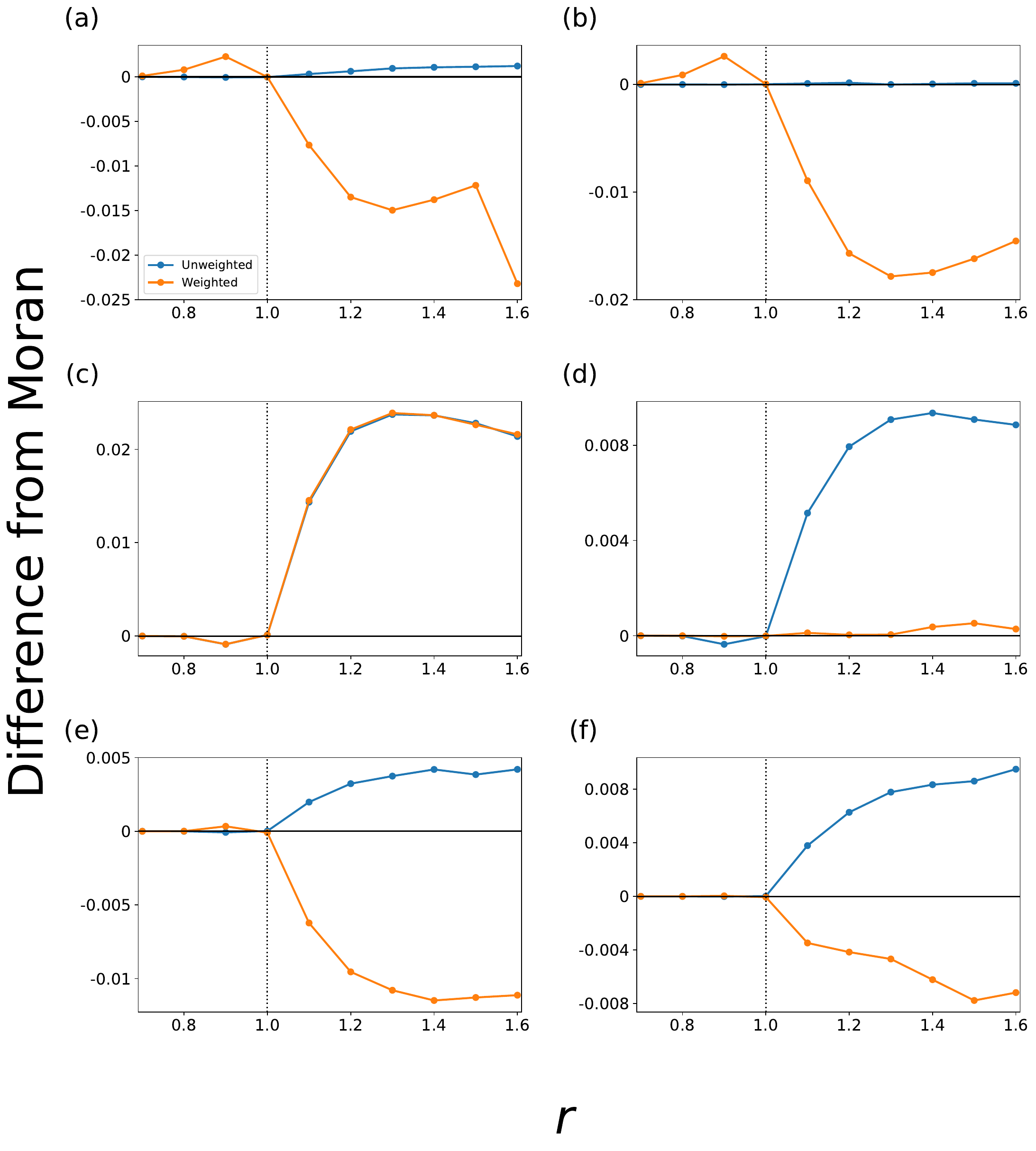}
	\caption{Fixation probability of the empirical networks. (a) Raccoons. (b) Primates. (c) Ants. (d) Sparrows. (e) Kilifi. (f) Hospital. The dotted lines represent $r=1$. 
	}
	\label{fig:empirical_figures}
\end{center}
\end{figure}

We show in Fig.~\ref{fig:empirical_figures} the fixation probability of the unweighted and weighted empirical networks relative to the fixation probability of the Moran process, with one empirical network in each panel. Figure~\ref{fig:empirical_figures}(a) indicates that the unweighted raccoon network is a weak amplifier of selection, whereas the weighted raccoon network is a relatively strong suppressor of selection. The results for the primate networks, shown in Fig.~\ref{fig:empirical_figures}(b), are qualitatively the same as those for the raccoons networks shown in Fig.~\ref{fig:empirical_figures}(a). Figure~\ref{fig:empirical_figures}(c) shows the results for ants' colony networks. Different from the other networks, this network shows almost indistinguishably similar fixation probability as a function of $r$ between the unweighted and weighted variants of the network, both of which are amplifiers of selection. Figure~\ref{fig:empirical_figures}(d) indicates that the unweighted sparrow network is a moderately strong amplifier of selection and that the weighted sparrow network is a much weaker amplifier of selection than the unweighted counterpart.  Figure~\ref{fig:empirical_figures}(e) indicates that the unweighted Kilifi network is an amplifier of selection and that the weighted version is a suppressor of selection.  Lastly, the results for the hospital networks, shown in Fig.~\ref{fig:empirical_figures}(f), are similar to those for the Kilifi networks (see Fig.~\ref{fig:empirical_figures}(e)).

 Overall, these results indicate that weighted networks tend to make the network less amplifying than their unweighted counterparts, with an exception shown in Fig.~\ref{fig:empirical_figures}(c).
	
\section{Discussion}

We have shown that, under the Bd updating rule and uniform initialization, a large proportion of the weighted networks on six nodes are less amplifying than their corresponding unweighted networks. Furthermore, a majority of these less amplifying networks are suppressors of selection relative to the Moran process. This result is in stark contrast to the case of unweighted networks, for which there are 100 amplifiers and just 1 suppressor of selection among the 112 possible networks with six nodes~\cite{alcalde2017suppressors}. In the case of the dB rule and uniform initialization, unweighted networks are never an amplifier of selection~\cite{tkadlec2020limits}, and most unweighted networks are suppressors of selection~\cite{hindersin2015most,alcalde2017suppressors}. Weighted networks on six nodes under the same condition did not yield any amplifiers of selection. Furthermore, under the dB rule, a majority of the weighted networks on six nodes were stronger suppressors of selection than the corresponding unweighted networks, which are already suppressors of selection in a majority of cases. The result that many weighted networks are suppressors of selection or at least less amplifying than their unweighted counterparts also holds true for five out of the six empirical networks we have investigated. For the other empirical network, the weighted and unweighted networks behaved similarly in terms of the fixation probability. For symmetric networks, which allow semi-analytical solutions, the results are more nuanced. We have shown that, depending on the value of parameters controlling some edge weights, the weighted complete graph and weighted star graph can be either more amplifying or less amplifying than their unweighted counterparts.  Overall, we conclude that introducing an edge weight is an easy method to make the network more suppressing than the original unweighted networks or the Moran process. The present results add weighted networks to an existing list of popular variations of the networks, such as the directed networks~\cite{masuda2009evolutionary}, temporal networks~\cite{Bhaumik2023switching}, hypergraphs~\cite{ruodan2023hypergraph}, and multilayer networks~\cite{ruodan2023multilayer}, that induce suppressors of selection under the Bd rule and uniform initialization, for which the amplifiers of selection are a norm for simple networks.
  
For evolutionary social dilemma games on networks, dynamics of edge weights with which profitable edges are strengthened and unprofitable edges are weakened promotes cooperation~\cite{chu2017coevolution, cao2011evolution}. Aspiration-based coevolution of edge weights~\cite{shen2018aspiration} and reputation-based adaptive adjustment of edge weights~\cite{guo2018reputation,li2019reputation} can also lead to evolution of cooperation. Even for static weighted networks, an increased heterogeneity in edge weights tends to enhance cooperation~\cite{ma2011heterogeneous,huang2015understanding,iwata2016heterogeneity}.
Inspired by these studies, dynamically varying edge weights in constant-selection evolutionary processes may yield interesting phenomena.

From engineering points of view, changing the edge weight may be easier than creating new edges or severing existing edges. Furthermore, weighted rather than unweighted networks better justify application of a perturbation theory that aims to assess the effect of changing edge weights by a small amount on properties of fixation such as the fixation probability and time. Note that, for evolutionary games on networks, such perturbation theory has been developed~\cite{amaral2020heterogeneity,amaral2020strategy,Lingqi2023perturbation}. Optimization of the edge weight given the unweighted network structure may also be an easier problem than the optimization of the network structure because only the latter is apparently a combinatorial problem. Construction of arbitrarily strong amplifiers of selection by weighted networks has already been examined \cite{pavlogiannis2018construction,tkadlec2021fast}. \add{Although one can engineer to create strong amplifiers, as these studies showed, an overall tendency found in the present study is that random assignment of edge weights tends to make networks suppressors of selection.}
%
%
Furthermore, megastars, which are directed unweighted networks, provide a family of strongly amplifying networks \cite{galanis2017amplifiers}. Megastars and the so-called dense incubators were later shown to be the most amplifying family of strongly amplifying networks among all the strongly connected directed networks and connected undirected networks, respectively \cite{goldberg2019asymptotically}. Maximization of the fixation probability with respect to the choice of the initially mutant nodes has also been examined
\cite{masuda2015opinion}. Another type of optimization problem in fixation dynamics is the positional Moran process~\cite{brendborg2022fixation}. In the positional Moran process, the fitness difference of the mutants is only realized on a subset of the entire node set, which one can engineer. It may be interesting to further explore optimization of weighted networks in terms of the fixation probability or time, or the extent of amplification or suppression of selection, including through perturbation theory.



\section*{Funding}\label{sec:funding}
 N.M. acknowledges support from the Japan Science and Technology Agency (JST) Moonshot R\&D (under Grant No. JPMJMS2021), the National Science Foundation (under Grant Nos. 2052720 and 2204936), and JSPS KAKENHI (under grant Nos. JP 21H04595 and
23H03414).
	

\appendix
\titleformat{\section}[block]{\normalfont\Large\bfseries}{\thesection}{1em}{}
\renewcommand{\thesection}{Appendix \Alph{section}}

\section{Transition probabilities for the weighted complete graph}\label{sec:appendix_A-transition-probabilities-for-the-weighted-complete-graph}

In this section, we describe the transition probabilities for the fixation dynamics on the weighted completed graph. As described in \del{Sec.}\add{section}~\ref{sec:completegraph}, we consider a complete weighted graph on $N$ nodes where the nodes are divided into two sets, one with $N_1$ nodes and the other with $N-N_1 \equiv N_2$ nodes. A state in the dynamics on this network is specified by the ordered pair $\left(i,j\right)$, where $i \in \{0, 1, \ldots N_1\}$ and $j \in \{ 0, 1, \ldots, N_2 \}$. State $\left(i,j\right)$ indicates that $i$ out of the $N_1$ nodes and $j$ out of the $N_2$ nodes are occupied by mutants.

The transition probability matrix  is given by
\begin{equation}
		T_{\left(i,j\right)\rightarrow\left(i',j'\right)} = \begin{cases}
			
				\frac{r i}{F_1}\cdot \frac{w_1 (N_1-i)}{s_1}+\frac{r j}{F_1}\cdot \frac{N_1-i}{s_2} & \text{if } i = 0 \text{ and } i'=1,
				\\[1mm]
					\frac{N_1-i}{F_1}\cdot\frac{w_1 i}{s_1}+\frac{N_2-j}{F_1}\cdot\frac{i}{s_2} & \text{if } i=1 \text{ and } i'=0,
				\\[1mm]
								\frac{r i}{F_1}\cdot\frac{N_2-j}{s_1}+\frac{r j}{F_1}\cdot\frac{w_2(N_2-j)}{s_2}
				& \text{if }j=0 \text{ and } j'=1,
				\\[1mm]
				\frac{N_1-i}{F_1}\cdot \frac{j}{s_1}+\frac{N_2-j}{F_1}\cdot \frac{w_2 j}{s_2} & \text{if } j=1 \text{ and }j'=0,
				\\[1mm]

				1-\sum\limits_{\mathclap{\substack{(i'',j'',k'')\neq \\ (i,j,k)}}}T^{\left(1\right)}_{(i,j,k) \rightarrow (i'',j'',k'')} & \text{if } (i',j')=(i, j),
				\\[1mm]
				0 & \text{ otherwise,}
			\end{cases}
			\label{eqn:weighted-complete-graph}
\end{equation}
where
\begin{align}
F_1 =& (i+j) r+(N-i-j),\\
s_1 =& w_1 (N_1-1)+N_2,\\
s_2 =& w_2 (N_2-1)+N_1.
\end{align}
We recall from Eq.~\eqref{eqn:recursive-fixation} that $x_{f(\bm{s})}$ is the probability that the mutant fixates when the evolutionary dynamics start from state $\bm{s}$. 
Using Eq.~\eqref{eq:rho-uniform-final}, we obtain
	\begin{equation}
		\label{eqn:fixn_probab_time_t_0}
		\rho = \frac{N_1}{N}  x^*_{f(\left(1,0\right))}+\frac{N_2}{N}x^*_{f(\left(0,1\right))}
	\end{equation}
because $\left(1,0\right)$ and $\left(0,1\right)$ are the only states in which just one mutant is present. Note that there are $N_1$ ways to realize state $\left(1,0\right)$, depending on which one node out of the $N_1$ nodes is of the mutant type, and $N_2$ ways to realize state $\left(0,1\right)$.

\section{Transition probabilities for the weighted star graph}\label{sec:appendix_B-transition-probabilities-for-the-weighted-star-graph}

In this section, we describe the transition probabilities for the fixation dynamics on the weighted star graph. A state in the dynamics on this network is specified by the ordered triplet $\left(i,j,k\right)$, where $i \in \{ 0, 1 \}$, $j \in \{ 0, 1, \ldots N_1 \}$, and $k \in \{ 0, 1, \ldots, N_2\}$. State $\left(i,j,k\right)$ indicates that
the hub node having degree $N-1$ hosts a resident (if $i=0$) or mutant (if $i=1$), $j$ out of the $N_1$ leaves are mutants, and $k$ out of the $N_2$ leaves are mutants.

The transition probability matrix is given by
\begin{equation}
	T_{\left(i,j,k\right)\rightarrow\left(i',j',k'\right)} = \begin{cases}
		\frac{r(j+k)}{F_2} & \text{if } i=0 \text{ and } i'=1,
		\\[1mm]
		\frac{1}{F_2}\cdot\frac{w j}{s_3} & \text{if } i = 0 \text{ and } j=j'+1,
		\\[1mm]
		\frac{1}{F_2}\cdot \frac{b}{s_3} & \text{if } i=0 \text{ and }k= k'+1,
		\\[1mm]
		\frac{(N_1-j)+(N_2-k)}{F_2} & \text{if }i=1 \text{ and }i' = 0,
		\\[1mm]
		 \frac{r}{F_2}\cdot \frac{w (N_1-j)}{s_3} & \text{if }i=1 \text{ and } j' = j+1,
		\\[1mm]
		\frac{r}{F_2}\cdot\frac{N_2-k}{s_3} & \text{if }i=1 \text{ and } k' = k+1,
		\\[1mm]
		1-\sum\limits_{\mathclap{\substack{(i'',j'',k'')\neq \\ (i,j,k)}}}T_{(i,j,k) \rightarrow (i'',j'',k'')} & \text{if } (i',j',k')=(i, j, k),
		\\[1mm]
		0 & \text{ otherwise,}
	\end{cases}
	\label{eqn:weighted-star-graph}
\end{equation}
where
\begin{equation}
	F_2=r(i+j+k)+\left[N_1+N_2+1-(i+j+k)\right]
\end{equation}
and
\begin{equation}
	s_3=w N_1+N_2.
\end{equation}
Then, we solve the system of linear equations given by Eq.~\eqref{eqn:recursive-fixation}.
Finally, using Eq.~\eqref{eq:rho-uniform-final}, we obtain
\begin{equation}
	\label{eqn:fixn_probab_time_t_0_star}
	\rho = \frac{1}{N}  x^*_{f(\left(1,0,0\right))}+\frac{N_1}{N}  x^*_{f(\left(0,1,0\right))}+\frac{N_2}{N}x^*_{f(\left(0,0,1\right))},
\end{equation}
because $(1,0,0)$, $(0,1,0)$, and $(0,0,1)$ are the only states in which one mutant is present. The hub node being initially occupied by a mutant corresponds to state $(1,0,0)$. The probability of this event is $1/N$. Similarly, there are $N_1$ ways in which the network is in state $(0,1,0)$ and $N_2$ ways in which the network is in state $(0,0,1)$.                    

\add{\section{Description of the empirical networks}}\label{sec:description-of-the-empirical-networks}

\add{In secion~\ref{sec:Empirical-networks}, we used the following six empirical networks, which we acquired from \url{https://networkrepository.com/}~\cite{nr-aaai15}.} 

\add{First, we use a social network of
	raccoons}~\cite{reynolds2015raccoon}. \add{The data were collected using proximity logging collars on a wild suburban raccoon population within a 20-ha area of Ned Brown Forest Preserve in Cook County, Illinois, USA, observed for $52$ weeks from July 2004 to June 2005. An event was recorded whenever two raccoons came in close proximity (i.e., 1-1.5m) of each other. A node represents a raccoon. An edge represents a proximity event between two raccoons. We set the edge weight to the total time that the given pair of raccoons spent in proximity. In total, there are $N=24$ nodes and $1997$ edges.}

\add{Second, we use a primate network based on a wild group of $N=25$ \textit{Macaca fuscata} individuals in Yakushima, Japan} \cite{griffin2012community}\add{. The population was free ranging and not captive. An edge represents grooming interaction between the individuals. There are 1340 edges. An event implies an interaction for greater than one minute. The edge weight is equal to the number of grooming interactions between the two individuals.}

\add{Third, we use data from an ants' colony}~\cite{ants2015quevillonsocial}\add{. Each of the $N=39$ nodes represents an ant in a colony. Each of the 330 edges represents a trophallaxis event, which was recorded when the two ants were engaged in mandible-to-mandible contact for greater than one second. The edge weight is equal to the number of events involving the two ants.}

\add{Fourth, we use a social network of sparrows }\cite{arnberg2015social}\add{. A flock was defined as a group of birds within an approximately 5-meter radius. A node represents a bird. Edges represent the so-called simple ratio association index}~\cite{cairns1987comparison} \add{distributed between 0 and 1, encoding flock co-membership. There are $N=40$ nodes and 305 edges.}

\add{Fifth, we use contacts between members of five households in the Matsangoni sub-location within the Kilifi Health and Demographic Surveillance Site (KHDSS) in coastal Kenya}~\cite{kilifi2016quantifying}\add{. A household was characterized as a collection of individuals who shared a common kitchen. Each participant in the study wore a sensor capable of detecting another sensor within a 1.5-meter radius. Each node represents a household member. An edge denotes a recorded interaction between two members. The study documented 47 nodes and 219 pairwise interactions across individuals from different households and 32,426 pairwise interactions within the same household. The edge weight is equal to the number of interaction events between two individuals.}

\add{Sixth, we use a human contact network in a hospital}~\cite{hospitalvanhems2013estimating}\add{. Data collection took place in a university hospital's geriatric unit in Lyon, France, between December 6, 2010, at 1 pm, and December 10, 2010, at 2 pm. Nineteen beds were located in the unit. Thirty-one patients were hospitalized, and 50 professionals worked in the unit during the recording period. Among a total of 81 individuals, $N=75$ individuals, i.e., 29 patients and 46 medical staff, participated in the study. Among the medical staff were $27$ nurses or nurses’ aides, $11$ medical doctors, and $8$ administrative staff members. An edge represents a time-stamped contact between two individuals; there are $32,424$ time-stamped edges. The edge weight is the number of times an event occurred between a pair of nodes.}

\bibliographystyle{unsrt}

\end{document}